\documentclass[apj]{emulateapj}

\shorttitle{Eigen-Reconstruction of Scalar Perturbations}
\shortauthors{M. S.~ Esmaeilian, M.~ Farhang, SH.~ Khodabakhshi}

\usepackage{lineno}
\usepackage{color, soul}
\usepackage[colorlinks=true,
            linkcolor=black,
            urlcolor=black,
            citecolor=blue, breaklinks=true]{hyperref}
\usepackage{url}
\usepackage{amsmath}
\usepackage{amssymb}
\usepackage{gensymb}
\usepackage{bm}

\usepackage{graphics}
\usepackage{graphicx}
\usepackage{natbib}
\usepackage{psfig}
\input{epsf}
\usepackage{ae,aecompl}
\usepackage{booktabs}

\newcommand{\be}{\begin{equation}}
\newcommand{\ee}{\end{equation}}

\begin{document}
\title{Detectable data-driven features in the  primordial scalar power spectrum}

\author{Muhammad Sadegh Esmaeilian\altaffilmark{1}, Marzieh Farhang\altaffilmark{1}, Shirin Khodabakhshi\altaffilmark{2}}
\affil{\scriptsize
$^{1}${Department of Physics, Shahid Beheshti University, 1983969411,  Tehran Iran}\\
$^{2}${Department of Physics, University of Tehran, Tehran 14395, Iran}}

\email{m_farhang@sbu.ac.ir}

\begin{abstract}
In this work we explore the power of  future large-scale surveys to constrain possible deviations from the standard single-field slow-roll inflationary scenario. 
Specifically, we parametrize possible fluctuations around the almost scale-invariant primordial scalar power spectrum in a model independent way. 
We then use their imprints on the simulated matter distribution, as observed by the galaxy clustering and weak lensing probes of Euclid and Square Kilometer Array, to construct the best constrainable patterns of fluctuations. 
For comparison, we make similar forecasts for a futuristic CMB-S4-like survey. 
The modes are found to have similar, yet shifted, patterns, with increasing number of wiggles as the mode number increases. The forecasted constraints are tightest for CMB anisotropies and galaxy clustering, depending on the  details of the specifications of the survey. 
As case studies, we explore how two greatly different physically motivated patterns of primordial power spectrum are reconstructed by the proposed modes.
We propose a figure of merit based on the amount of  information delivered by the modes to truncate the mode hierarchy which is automatically generated by the analysis.
\end{abstract}

%------------------------
\section{Introduction}
%------------------------
Future cosmological experiments targeting the largest scales will map the distribution of matter in the Universe at various redshifts on a large span of cosmological scales. 
The $3$D maps will reveal unprecedented amount of information about the evolution of the inhomogeneities and the formation of structures. 
This capacity is extensively explored through studying different cosmological and gravitational extensions to the standard model of $\Lambda$CDM, mainly in the late universe \citep[see, e.g., ][]{Cardone_2013,Raccanelli_2016,Bull_2016,casas2017linear,Copeland_2018,Copeland_2020,Apa_Res_2020}. 

Future surveys will also enhance the window to the physics of the early Universe. 
The single field slow-roll models of inflation predict nearly-Gaussian and almost scale--invariant power spectrum for scalar and tensor perturbations \citep{Starobinsky1979,1982PhLB..108..389L}, minimally described by the amplitudes $A_{\rm s(t)}$  and spectral indices $n_{\rm s(t)}$ with ${\rm s(t)}$ representing the scalar (tensor) fluctuations. These parameters are currently measured by the high precision observations of the cosmic microwave background (CMB) by the {\it Planck} satellite \citep{pl18,pl18inf}. However, there is still observational room for small deviations from this prediction.
Several authors have investigated the sensitivity of  future measurements of  matter distribution to the physics of early Universe and estimated the detectability of the traces left by various early universe scenarios on the matter distribution \citep[see, e.g., ][]{Fedeli_2011,Huang_2012,Fedeli_2012,Huillier18,Ballardini_2019,Resco_2020,Debono_2020}.

In this work we also aim to exploit the information in the future large scale surveys to constrain deviations from the standard inflationary paradigm. 
Specifically, we follow a model--independent  parametrization in modeling  possible perturbations to the scale-invariant power spectrum, and avoid theoretical biases. 
In our blind analysis we start by introducing perturbations in the primordial power spectrum in a semi-blind way and use their traces in the simulated data to construct the most tightly constrainable patterns of fluctuations. 
These modes are not driven by any physically motivated models, except for being expanded around an unavoidable assumption for the fiducial, not surprisingly taken to be the almost scale-invariant inflationary power spectrum.  Our focus will be the scalar perturbations as they are by far the only observable seed to produce matter inhomogeneities.  
The  model--independent parametrization of possible fluctuations  around a fiducial model has already been used in different works in cosmology \cite[see, e.g., ][]{Zhao_2009,Ishida_2011,Hall_2013,Sapone_2014,Regan_2015,Liu_2016,Feng_2016,Huang_2017,Taylor_2018,Farhang_2019,sharma2020reconstruction}. 

 The  best observable patterns or eigenmodes generated this way can be used in the availability of data to reconstruct possible deviations in the power spectrum and to
 assess the assumptions behind the scenarios  of early Universe. 
On the other hand,  the predicted power spectrum of any extensions to the standard inflationary paradigm can be projected on the eigenmodes. A large projection onto the tightly constrainable modes 
 implies the detectability of the imprints of the scenario. Otherwise, theoretical priors are highly required to push in favor of that model.  
 
In this work we use the galaxy clustering and weak lensing probes of the Euclid\footnote{https://www.euclid-ec.org} \citep{Euclid:2011,Amendola_2013} and SKA\footnote{https://www.skatelescope.org} \citep{Santos:2015} surveys as the main datasets. We also use simulations for the CMB power spectrum of the CMB-S4\footnote{https://www.cmb-s4.org} futuristic mission \citep{abazajian2016cmbs4,abitbol2017cmbs4} for comparison with the data in the matter distribution. The fiducial model corresponds to the standard inflationary $\Lambda$CDM scenario with parameters consistent with the best-fit measurements of {\it Planck} dataset \citep{pl18}.
The rest of this paper is organized as follows. In Section~\ref{sec:method} we introduce the parameterization used to quantify deviations around the primordial power spectrum, and review the mathematical framework for the mode construction from these parametrization. 
The probes and their corresponding Fisher matrices are discussed in Section~\ref{sec:prob}. Section~\ref{sec:results} presents the modes and assesses their hierarchical importance by introducing a figure of merit based on the parameter space spanned by the modes. 
In this section we also investigate the ability of the modes in the reconstruction of two different models of power spectrum with different sets of simulated datasets.
We conclude with a discussion in Section~\ref{sec:discussion}.

%------------------------
\section{Eigen-reconstruction of perturbations}\label{sec:method}

In this section we investigate the sensitivity of several cosmological probes to features in the primordial scalar power spectrum, or PSPS afterwards. We parametrize these features as amplitudes of fluctuations, expanded in orthonormal sets of basis functions, around the nearly scale--invariant power spectrum which is predicted by single--field slow--roll inflationary models \citep{Starobinsky1979,1982PhLB..108..389L}. We exploit the information in these probes to look for traces left by possible deviation from the scale-invariant spectrum through a semi-blind model-independent approach. 

%--------------------------------------------------------------------------------------
  \begin{figure*}
   \begin{center}
  \includegraphics[scale=0.43]{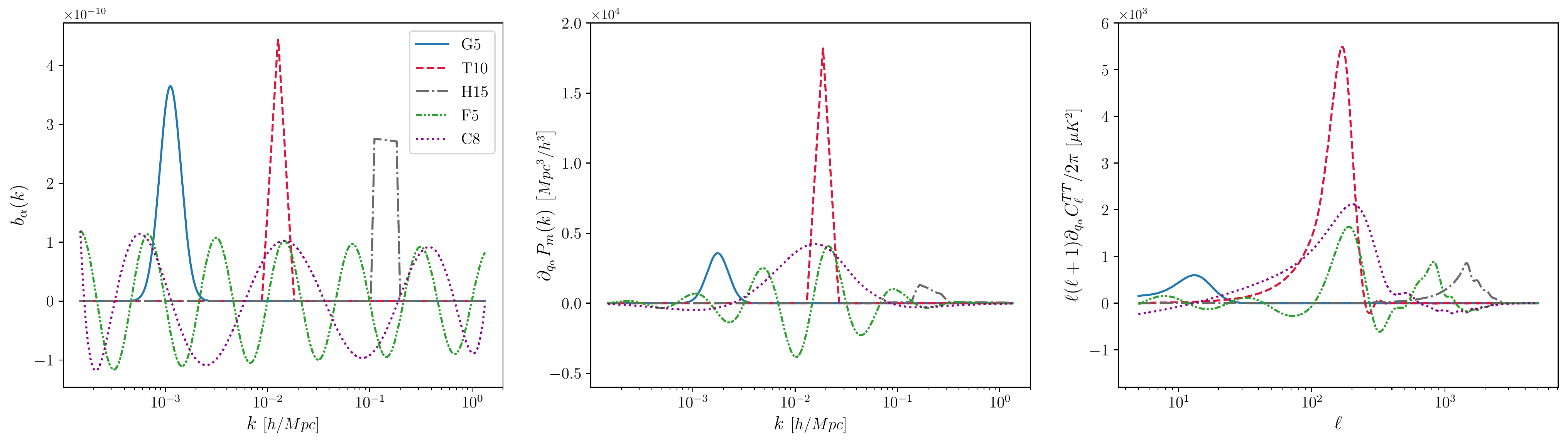}
     \caption{Left: Localized (labeled as G for Gaussian, T for triangular and H for top-hat) and non-localized (labeled as F for Fourier series and C for Chebyshev polynomials) basis functions, with $N=20$. The numbers in the curve labels correspond to the number of basis function. Middle and right: The response of the matter and CMB temperature power spectra to perturbations in the PSPS parametrized by $q_i$, i.e., the amplitudes of the basis functions of the left panel. }
        \label{fig:base}
   \end{center}
   \end{figure*}
%--------------------------------------------------------------------------------------
We express the perturbed PSPS as 
\begin{equation}
  \mathcal{P}(k) =  \mathcal{P}_0(k)  \left[1 + \delta_{\mathcal{P}}(k)\right]
  \label{eq:psps} 
\end{equation}

where $\mathcal{P}_0(k)=A_{\rm s}(k \slash k_{*})^{n_{\rm s}-1}$, and $A_{\rm s}$ and $n_{\rm s}$ are the amplitude and spectral tilt of the PSPS, calculated at the scalar pivot wavenumber $k_{*}=0.05 {\rm Mpc^{-1}}$. 

Theoretically, any set of complete orthogonal basis functions, $\{b_{\alpha}(k)\}$ where $\alpha=1, ..., N$,  can be used to expand the perturbations, 

\begin{equation}
\delta_{\mathcal{P}}(k) =\sum_{\alpha=1}^{N} \eta_\alpha\,b_{\alpha}(k) \label{eq:base}
\end{equation}
 where $N$ is the number of the basis functions, and in principle $N\rightarrow \infty$. In practice, however, a finite set of basis functions suffices to span the space of physically interesting perturbations.
 
In the rest of this section we briefly review the basis functions used in this work for the expansion of fluctuations (Section~\ref{sec:base}), and  the method of principal component analysis to construct the best-constrained modes of perturbations (Section~\ref{sec:pca}).    %----------------------------------------------------------------------------
  \subsection{Basis Functions}\label{sec:base}
We explore various sets of local and nonlocal basis functions for the expansion of perturbations to the scale-invariant PSPS.
As the local sets we use the Gaussians  

\begin{equation}
b_{\alpha}^{\rm G}(k) \propto \exp\left[-\frac{(y - y_\alpha)^{2}}{2 \sigma_{\rm G}^{2}}\right],
\end{equation}
 the triangulars
 
\begin{equation}
  b_\alpha^{\rm T}(k) \propto \left\{\begin{array}{ll} 1-\left|y-y_\alpha\right| / \sigma_{\rm T} \:\:\:& \text { if }\left|y-y_\alpha\right|<\sigma_{\rm T} \\ 0 \:\:\: & \text { otherwise }\end{array}\right.
  \end{equation}
and top-hats, 

 \begin{equation}
  b_\alpha^{\rm H}(k) \propto \left\{\begin{array}{ll} 1 \:\:\:& \text { if }\left|y-y_\alpha\right|<\sigma_{\rm H} \\ 0 \:\:\: & \text { otherwise. }\end{array}\right.
  \end{equation}
 where $y=\ln k$ and 
 $y_\alpha$ and $\sigma_{\rm X}$ (with ${\rm X=T,G,H}$)  represent  the bin centers and widths respectively.
 We choose $\sigma_{\rm{G,T}}=\Delta_{\ln k}/N$ and $\sigma_{H}=1.5\Delta_{\ln k}/N$ where $N$ is the number of bins and $\Delta_{\ln k}=\ln(k_{\rm max}/k_{\rm min})$ is the width of the $k$--range of interest with $k_{\rm min}$ and $k_{\rm max}$ labeling the minimum and maximum $k$ used in the analysis.   
 
As the non-local, or extended, bases, we choose the Fourier series

  \begin{equation}
  b_{\alpha}^{\rm F}(k) \propto \left\{\begin{array}{ll} \sin\,(\alpha \pi y) & \text { } \alpha=1,2, ... \\ \cos(\alpha\pi y) & \text { } \alpha =0,1,2, ...\end{array}\right.
  \end{equation}
  where $ y = \ln (k/k_{\text{mid}}) \slash \Delta_{\ln k}$ and $k_{\text{mid}}= (k_{\text{min}}+k_{\text{max}})\slash 2$, and the Chebyshev polynomials of the first kind, constructed from the recursion formula

  \begin{equation}
  b_{\alpha}^{\rm C}(k) \propto \left\{\begin{array}{ll} y^\alpha & \text { if } \alpha = 0, 1 \\ 2yb_{\alpha-1}^{\rm C}(k) - b_{\alpha-2}^{\rm C}(k) & \text { if } \alpha \geqslant 2\end{array}\right.
  \end{equation}

   The basis functions are normalized with respect to the appropriate weight function $w(k)$, 

  \begin{equation}
  \int_{k_{\min }}^{k_{\max }} b_{\alpha}(k) b_{\beta}(k) w(k) \mathrm{d} \ln k=\delta_{\alpha \beta}.
  \end{equation}
We have $w(k)=1$ for all the bases used in this work except for the Chebyshev polynomials where $w(k) = 1 \slash \sqrt{1-k^{2}}$. 
  
  Figure~\ref{fig:base} shows the perturbed PSPS and the corresponding responses in the matter and CMB power spectra for different local and nonlocal basis functions, with $k_{\rm min}=10^{-4}h/{\rm Mpc}$, $k_{\rm max}=1h/{\rm Mpc}$ and $N=20$.   
  
   %--------------------------------------------------------------------------------------
  \subsection{Principle Component Analysis}\label{sec:pca}
   %--------------------------------------------------------------------------------------
 In this section we explain how to use the observable imprints in the data, left by the perturbations to the scale-invariant PSPS, to construct new basis functions, or principal components, predicted to be best detectable by data.
  This method of principal component analysis, or PCA for short, by construction, extracts features in the parameter space of perturbations that data is most sensitive to.
  The data-driven modes benefit from being model-independent and serve as unbiased parametrization of deviation from the standard scenario in the absence of theoretical priors and privileges.  
  
Given a set of parameters $\bm{q}=\{q_\alpha\}$ where $\alpha=1,..,N $, the principal modes are constructed from the eigen vectors of the Fisher information matrix for $\bm{q}$,

  \begin{equation}
  [\mathcal{F}(\bm{q})]_{\alpha\beta}=-\left< \frac{\partial^{2} \ln P_{\rm f}}{\partial q_{\alpha} \partial q_{\beta}} \right>.
  \label{fish1}
  \end{equation}
 Here $P_{\rm f}\equiv P_{\rm f}(\bm{q}|d)= \mathcal{L}(d |\bm{q}) P_{\rm i}/\mathcal{E}$ is the Bayesian posterior distribution of the parameter set $\bm{q}$ for the dataset $d$, with $\mathcal{L}$, $P_{\rm i}$ and $\mathcal{E}$ respectively representing the likelihood of the data for the assumed set of parameters, the parameter distribution prior to the availability of $d$, and the evidence that serves as the normalization factor. 
 Assuming uniform priors for the parameters reduces the Fisher matrix to
  \begin{equation}
  [\mathcal{F}(\bm{q})]_{\alpha\beta}=-\left<\frac{\partial^{2} \ln \mathcal{L}}{\partial q_{\alpha} \partial q_{\beta}}  \right>.
  \label{fish2}
  \end{equation}
  The Fisher matrix can be eigen-decomposed as 
  \begin{equation}
  \mathcal{F} = X^{T} \Lambda\,X
  \end{equation}
  where $\Lambda$ is a diagonal matrix containing the Fisher eigenvalues and $X$ is a matrix whose columns represent the corresponding Fisher eigenvectors. 
  The eigenmodes are constructed from these eigenvectors

  \begin{equation}
  \mathcal{E}_{\alpha}(k) = \sum_{\beta = 1}^{N} X_{\alpha\beta}\,b_{\beta}(k)\, \sqrt{w(k)}
  \end{equation}
 where  $b_{\beta}$ and $w$  are the basis and weight functions.
The corresponding eigenvalue for each eigenvector gives the estimated error for the eigenmodes through $ \sigma_\alpha^2 =  \rm{diag} (\Lambda ^{-1})_\alpha$.
The eigenmodes $\mathcal{E}_\alpha(k)$ can be used for the expansion of any perturbations to the PSPS in the very same way as the local and nonlocal basis functions we started with (Equation~\ref{eq:base}).

In the next section we introduce the simulated datasets used in this work and the construction of the corresponding Fisher matrices.
%----------------------------------------
\section{Probes and Analysis}\label{sec:prob}
%----------------------------------------
In this work we use simulations for two sets of cosmological datasets, the power spectra of anisotropies in the temperature and $E$--mode polarization of CMB, and forecasts for the large scale distribution of matter.
Whenever required, the fiducial $\Lambda$CDM cosmology is considered  consistent with the best-fit parameters measured by the {\it Planck} dataset \citep{pl18}. 
We use the publicly available Boltzmann code \text{CAMB}\footnote{https://camb.info} \citep{Lewis:1999bs} for the calculation of the CMB and matter power spectra. 
The simulated datasets and their corresponding Fisher analysis are presented in Sections~\ref{sec:cmb} and \ref{sec:lss} respectively.
  \begin{table}
  \centering  
  \begin{tabular}{c|cccc}
  \noalign{\smallskip}
  \hline
  \noalign{\smallskip}
  Frequency &  NET &  FWHM  &  MD ($\%5$ SC)  &  MD ($\%50$ SC)  \\
   $(\mathrm{GHz})$ &  $\left(\mu {\rm K}\sqrt{\mathrm{s}}\right)$ & $(\mathrm{ arcmin})$ &  ($\mu {\rm K}\text{ arcmin}$) & ($\mu {\rm K}\text{ arcmin}$) \\

  \noalign{\smallskip}
  \hline \hline
  \noalign{\smallskip}
  $95$ & $1.5$ & $2.3$ & $1.3$ & $4.1$\\
  $150$ & $1.7$ & $1.5$ & $1.5$ & $4.6$\\
  \noalign{\smallskip}
  \hline
  \end{tabular}
  \caption{Future CMB-S4 specifications for the noise of CMB power spectra for the optimum homogeneous configuration with $6$m telescope aperture diameter \citep{Barron_2018}. The listed frequencies are the total available frequency channels in the Middle Band.  MD in the table stands for "Max Depth" and SC stands for "Sky Coverage".}
  \label{s4table}
  \end{table}
  %
%------------------------
 \subsection{Cosmic Microwave Background}\label{sec:cmb}
 %------------------------
 Here we investigate how the anisotropies in the temperature and $E$--mode polarization of the CMB can constrain features in the PSPS in a model-independent way.
A similar study on the primordial tensor perturbations, constrained by their imprints on the CMB $B$-mode anisotropies, was done in \cite{Farhang_2019}.

\begin{figure}
  \begin{center}
  \includegraphics[scale=0.6]{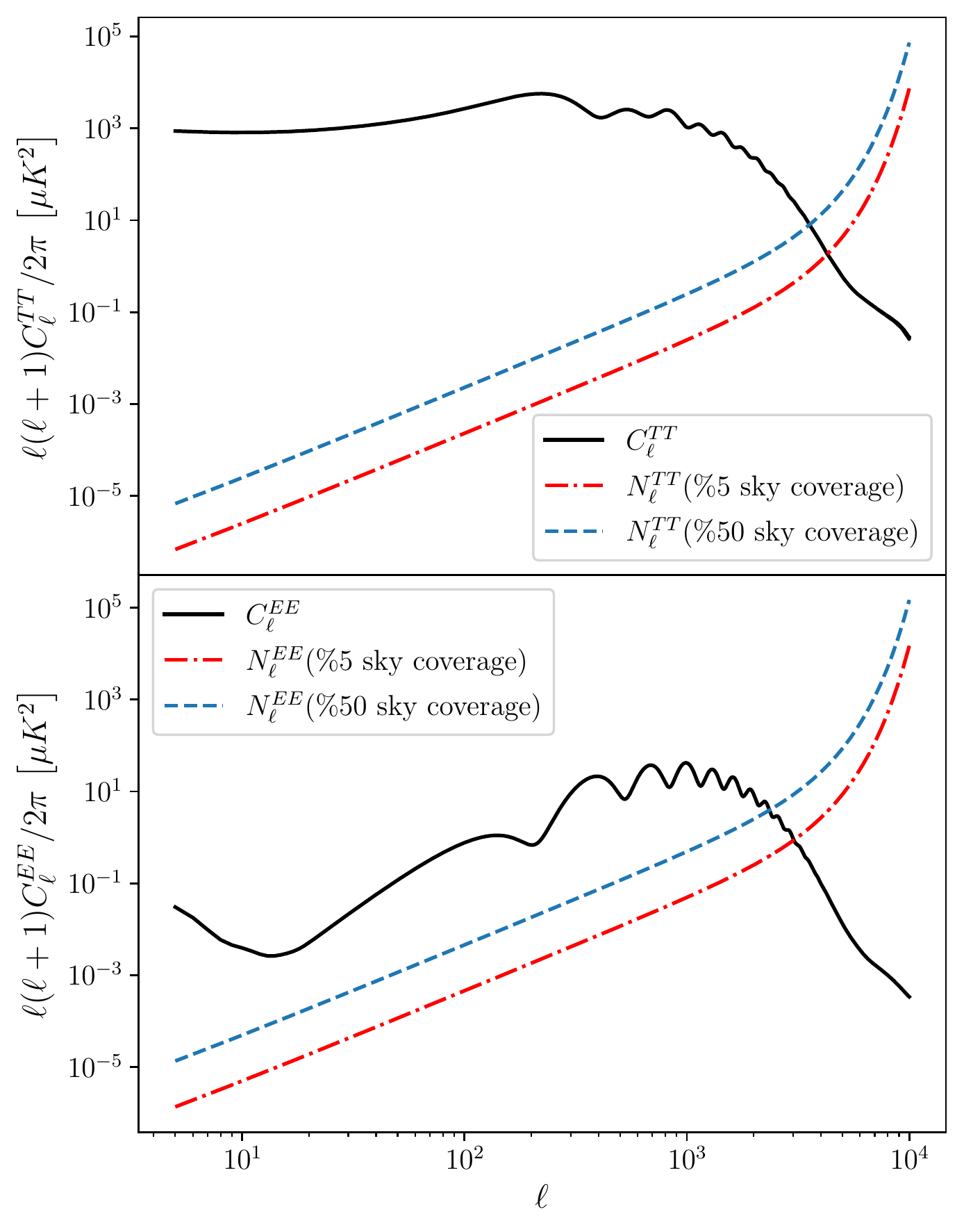}
  \caption{The CMB temperature and $E$--mode polarization power spectra versus the anticipated noise of a future CMB-S4-like experiment. The noise levels correspond to observation of $5\%$ and $50\%$ of the sky. }
    \label{fig:CMBNS4}
  \end{center}
  \end{figure}

 The Fisher matrix for the power spectra of CMB temperature and $E$--mode polarization is given by \citep[see. e.g.][]{Tegmark-1997CMB}
 
 %], \citep{Smith_2009}})
  %
  \begin{equation}
  F_{\alpha\beta} = f_{\text{sky}}\, \sum_{\ell=\ell_{\text{min}}}^{\ell_{\text{max}}} \frac{(2\ell +1)}{2} \operatorname{Tr} \left[\boldsymbol{C}_{\ell}^{-1} \frac{\partial \boldsymbol{C}_{\ell}}{\partial q_{\alpha}} \boldsymbol{C}_{\ell}^{-1} \frac{\partial \boldsymbol{C}_{\ell}}{\partial q_{\beta}}\right]
  \end{equation}
  where $q_{\alpha}$ and $q_{\beta}$ represent the cosmic parameters we are interested in, here the amplitudes of the perturbations.
   $\boldsymbol{C}_{\ell}$ is the covariance matrix   for multipole $\ell$,   %
   
  \begin{equation}
  \boldsymbol{C}_{\ell} \equiv\left[\begin{array}{cc} C_{\ell}^{T T}+N_{\ell}^{T T} & C_{\ell}^{T E} \\ C_{\ell}^{T E} & C_{\ell}^{E E}+N_{\ell}^{E E}\end{array}\right]
  \end{equation}
  where  $C_{\ell}^{XY}$ with $X, Y \in \{T,E\}$ represents the CMB power spectrum  for the CMB temperature, polarization and their cross-correlation respectively.
  We assume white Guassian instrumental noise with the power spectrum modeled by \citep{Knox_1995,Barron_2018}

  \begin{equation}
  N_{\ell}^{X X} = \frac{1}{\sum_{\nu} N_{\ell}^{XX,\nu}}
  \end{equation}
  where

  \begin{equation}
  N_{\ell}^{X X, \nu} = w_{X, \nu} \exp \left[-\ell(\ell+1) \frac{\theta_{\text{FHWM}, \nu}^{2}}{8\ln2}\right].
  \end{equation}
  Here  $\nu$ labels the frequency channel of the observation,  $\theta_{\text{FWHM}}$ is the full-width at the half-maximum  of the beam, and as before, $X \in \{T,E\}$.
   The weight function  $w_{X, \nu}$  is determined by the sensitivity of the instrument 
   and we assume $w_{E, \nu}=\sqrt{2} w_{T, \nu}$. 
  For the experimental setup, we use the specifications of the future CMB-S4 survey \citep{abazajian2016cmbs4,abitbol2017cmbs4}.
 Table~\ref{s4table} summarizes the specifications  for the two frequency channels and the different sky coverages used in this work. 
  Figure~\ref{fig:CMBNS4} compares the power spectra of CMB temperature and $E$-mode polarization with the expected noise spectra of a CMB-S4-like experiment.
 
%------------------------
\subsection{Large Scale Structure}\label{sec:lss}
%--------------------------------------------------------------------------------------
In this work we use simulations of the two future large scale surveys, the European Space Agency's Euclid \citep{Euclid:2011} and the Square Kilometer Array, or  the SKA mission with the two phases of SKA1 and SKA2 \citep{Santos:2015}, to investigate the detectability of data-driven features in the primordial scalar power spectrum. 
In the following we briefly review the two main probes of galaxy clustering  (Section~\ref{sec:gal}) and weak lensing (Section~\ref{sec:weak}).

  %-------------------------------------------------------------------------------------
    \subsubsection{Galaxy Clustering}\label{sec:gal}
  %-------------------------------------------------------------------------------------
  
    The galaxy clustering, or GC for short, is often characterized by the two point correlation function of galaxy number density, or equivalently by the galaxy power spectrum.  
    Higher order correlations are informative when nonlinear evolution of structures or non-Gaussian initial conditions are of interest. 
    The GC power spectrum is based on  the matter power spectrum $P_{\rm m}$ and is modeled as \citep{Seo:2007}

    \begin{equation}
    \begin{aligned} P_{\rm g}(k, \mu, z) = & \,A(z)\,b^{2}(z)\,(1+\beta_{\rm d}(z)\mu^{2})^{2}\,P_{\rm m}(k, z) \\ & \times 
    \label{eq:gc} e^{-k^{2}\mu^{2}(\sigma_{r}^{2} + \sigma_{v}^{2})}
    \end{aligned}
    \end{equation}
where $\mu$ is the cosine of the angle between the wave-vector $\vec{k}$ and the line of sight. The function $\beta_{\rm d}(z)$ accounts for the impact of the redshift-space distortion, modeled as 
$\beta_{\rm d}(z) = f(z) \slash b(z)$ in the linear theory, where $f(z)={\rm d}\ln\,G \slash {\rm d}\ln\,a$ and $G(z)$ is the growth function. 
  Also, $A(z)$ encodes the Alcock-Paczinsky effect \citep{1979Natur.281..358A}

    \begin{equation}
    A(z) = \frac{D_{A,f}^{2}(z)H(z)}{D_{A}^{2}(z)H_{f}(z)}
    \end{equation}
where the subscript ${\rm f}$ denotes the fiducial value of the parameter, $D_{A}(z)$ is the angular diameter distance to redshift $z$ and $H(z)$ is the Hubble parameter at $z$.
The exponential factor in Equation~\ref{eq:gc} suppresses the power spectrum due to uncertainties induced by spectroscopic measurements, $\sigma_{r} = \sigma_{z} \slash H(z)$, and the dispersion of pairwise peculiar velocities $\sigma_{v}$, chosen to be $300$ Km/s \citep{de_la_Torre_2012}. 
    \begin{table}
    \centering  
    \begin{tabular}{c|cccc}
    \noalign{\smallskip}
    \hline
    \noalign{\smallskip}
     & Euclid & SKA1 & SKA2  \\
    \noalign{\smallskip}
    \hline\hline
    \noalign{\smallskip}
    $A_{\text {sur} }$ & $15000$ & $5000$ & $30000$ & Survey area in the sky ($\text{deg}^{2}$)\\
    $\sigma_{z}$ & $0.001$ & $0.0001$ & $0.0001$ & Spectroscopic redshift error\\
    $z_{\text {min}}$ & $0.65$ & $0.05$ & $0.15$ & Min. limit for redshift bins\\
    $z_{\text {max}}$ & $2.05$ & $0.55$ & $2.05$ & Max. limit for redshift bins\\
    $\Delta_{z}$ & 0.1 & 0.1 & 0.1 & Redshift bin width\\
    \noalign{\smallskip}
    \hline
    \end{tabular}
    \caption{Specifications of the GC probe for the future surveys Euclid and SKA used in this work \citep{Amendola_2013,Harrison_2016,Sprenger_2019}.}
    \label{table:gc}
    \end{table}

 The Fisher matrix of GC for the parameter set $\bm{q}$, assuming  Gaussian distribution for the inhomogeneities, is given by \citep[see, e.g.,][]{Tegmark:1997}

    \begin{equation}
    \begin{aligned} F_{\alpha \beta} = & \frac{1}{8 \pi^{2}}\sum_{a} \int_{-1}^{+1} \mathrm{d} \mu \int_{k_{\text {low }}}^{k_{\text {up }}} \mathrm{d} k\,k^{2} \frac{\partial \ln P_{\text {g}}}{\partial q_{\alpha}} \\ & \times \frac{\partial \ln P_{\text {g}}}{\partial q_{\beta}}\left[\frac{n(z_{a}) P_{\text {g}}}{n(z_{a}) P_{\text {g}}+1}\right]^{2} V_{\text {sur }}(z_{a})
    \label{eq:fgc} 
    \end{aligned}
    \end{equation}
    where the sum is over the redshift bins of the experiment, $V_{\text {sur}}$ is the volume covered by survey and $n(z)$ is the observed galaxy number density. 
    The limits of the $k$--integral, i.e., $k_{\text {low}}$ and $k_{\text {up }}$, correspond to the minimum and maximum scales probed by the survey.
    The experimental specifications for the GC probes are presented in Table~\ref{table:gc}
    and the galaxy number density $n(z)$ for Euclid and SKA are taken from  \cite{Amendola_2013} and \cite{Bull_2016,Sprenger_2019} respectively. 
      
The nonlinear growth of structure plays an important role at late times and small scales. 
 As nonlinearities kick in, the mode coupling grows and the assumption of one-to-one correspondence in the $k$--space between the initial perturbations and the observed power spectrum breaks.
 On the other hand, the smaller scales get increasingly more informative for the measurement of most parameters as the number of available mode increases with increasing $k$. 
 This calls for an accurate nonlinear prescription used for solving the Boltzmann equations. 
The CAMB package used in this work is based on the fitting-scheme of Halofit to model the nonlinear regime \citep{halofit12}. Halofit is calibrated for $\Lambda$CDM and the standard inflationary PSPS and requires suitable extensions for other cosmological scenarios \citep{Smith:2018zcj,Ballardini:2019tuc}. 
 It is therefore extremely important that theoretical uncertainties in the modeling of the nonlinear regime be properly taken into account.   

Different approaches are proposed to account for nonlinear uncertainties when making forecast for the future large scale surveys \citep{Casas_2016,Sprenger_2019}. The most straightforward is to set a sharp cutoff on scales that can be accurately modeled and discard the modes beyond that.
However, the measurement of some parameters would be highly sensitive to the choice of cutoff,  making this sharp choice hard to justify. 
We instead choose a more smooth $k$--transition from the linear to nonlinear regime 
and model this transition by introducing a damping term in the Fisher matrix as

    \begin{equation}
    E(k) = \left\{\begin{array}{ll} 1 &  k_{\text{low}} \le k < k_{\text{cut}}
    \\ \exp(-\tau \left[\frac{k-k_{\text{cut}}}{k_{\text{cut}}}\right]^{2}) &   k_{\text{cut}} \le k \le k_{\text{up}}\end{array}\right.
    \end{equation}
The bounds $k_{\text{low}}$ and $k_{\text{up}}$ are the same as those in Equation~\ref{eq:fgc}.  
The lower limit of $k$ is set by the survey volume. As the cutoff scale we choose $k_{\text{cut}}=0.15~h/\text{Mpc}$ as the approximate scale where 
nonlinearity can no longer be ignored. 
For the upper bound we safely use $k_{\rm up}=0.25~h/\text{Mpc}$, as the 
damping term for the nonlinear uncertainty makes the 
contribution of the power spectrum beyond that negligible.
%--------------------------------------------------------------------------
    \subsubsection{Weak Lensing}\label{sec:weak}
%--------------------------------------------------------------------------
  %
    \begin{figure}
    \begin{center}
    \includegraphics[scale=0.6]{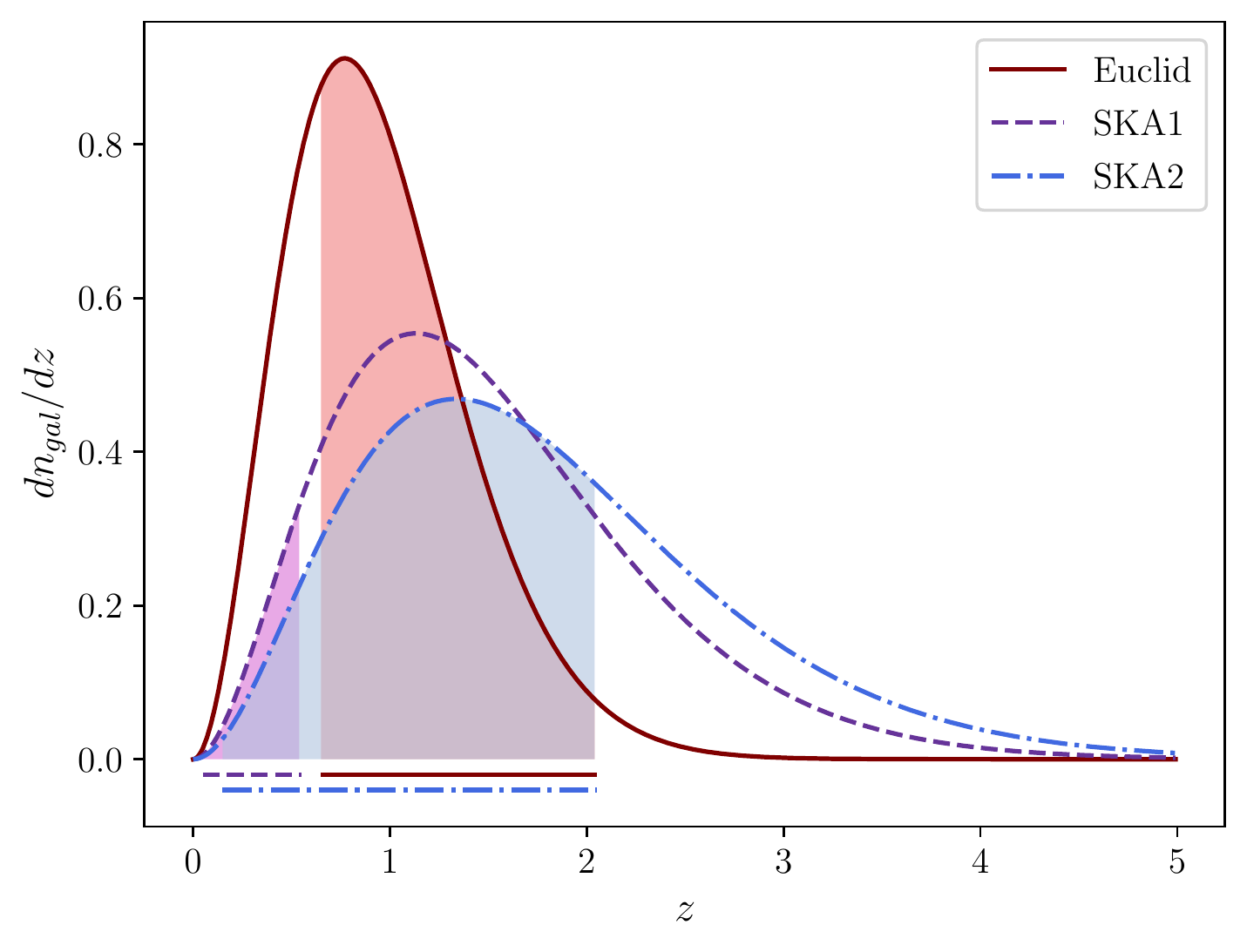}
    \caption{The normalized distribution function of the source redshifts  for the WL probes of Euclid and SKA surveys. The shaded regions correspond to the redshift ranges covered by the surveys.}
      \label{fig:NSRDF}
   \end{center}
   \end{figure}
    \begin{table}
    \centering 
    \begin{tabular}{c|cccc}
    \noalign{\smallskip}
    \hline
    \noalign{\smallskip}
     & Euclid & SKA1 & SKA2 &\\
    \noalign{\smallskip}
    \hline\hline
    \noalign{\smallskip}
    $f_{\text {\,sky} }$ & $0.3636$ & $0.1212$ & $0.7272$ & Sky coverage fraction\\
    $\sigma_{z}$ & $0.05$ & $0.05$ & $0.03$ & Photometric redshift error\\
    $\gamma_{\text{int}}$ & $0.22$ & $0.3$ & $0.3$ & Intrinsic galaxy ellipticity\\
    $n_{\theta}$ & $30$ & $10$ & $2.7$ & No. of galaxies per arcmin\\
    $z_{0}$ & $0.9$ & $1.0$ & $1.6$ & Median redshift over $\sqrt{2}$\\
    $\mathcal{N}_{\text{bin}}$ & $12$ & $12$ & $12$ & No. of tomographic bins\\
    $\alpha$ & $\sqrt{2}$ & $\sqrt{2}$ & $\sqrt{2}$ & Parameter used in Eq.~\ref{UNDR} \\
    $\beta$ & $2$ & $2$ & $2$ & Parameter used in Eq.~\ref{UNDR}\\
    $\eta$ & $1.5$ & $1.25$ & $1.25$ & Parameter used in Eq.~\ref{UNDR}\\
    $z_{\text {min}}$ & $0.65$ & $0.05$ & $0.15$ & Min. limit for redshift bins\\
    $z_{\text {max}}$ & $2.05$ & $0.85$ & $2.05$ & Max. limit for redshift bins\\
    \noalign{\smallskip}
    \hline
    \end{tabular}
    \caption{Specifications of the WL probe for the future surveys Euclid and SKA used in this work \citep{Amendola_2013,Harrison_2016,Sprenger_2019}.}
    \label{sp2table}  
    \end{table}
%-------------------------------------------------------
  \begin{figure}
  \begin{center}
  \includegraphics[scale=0.7]{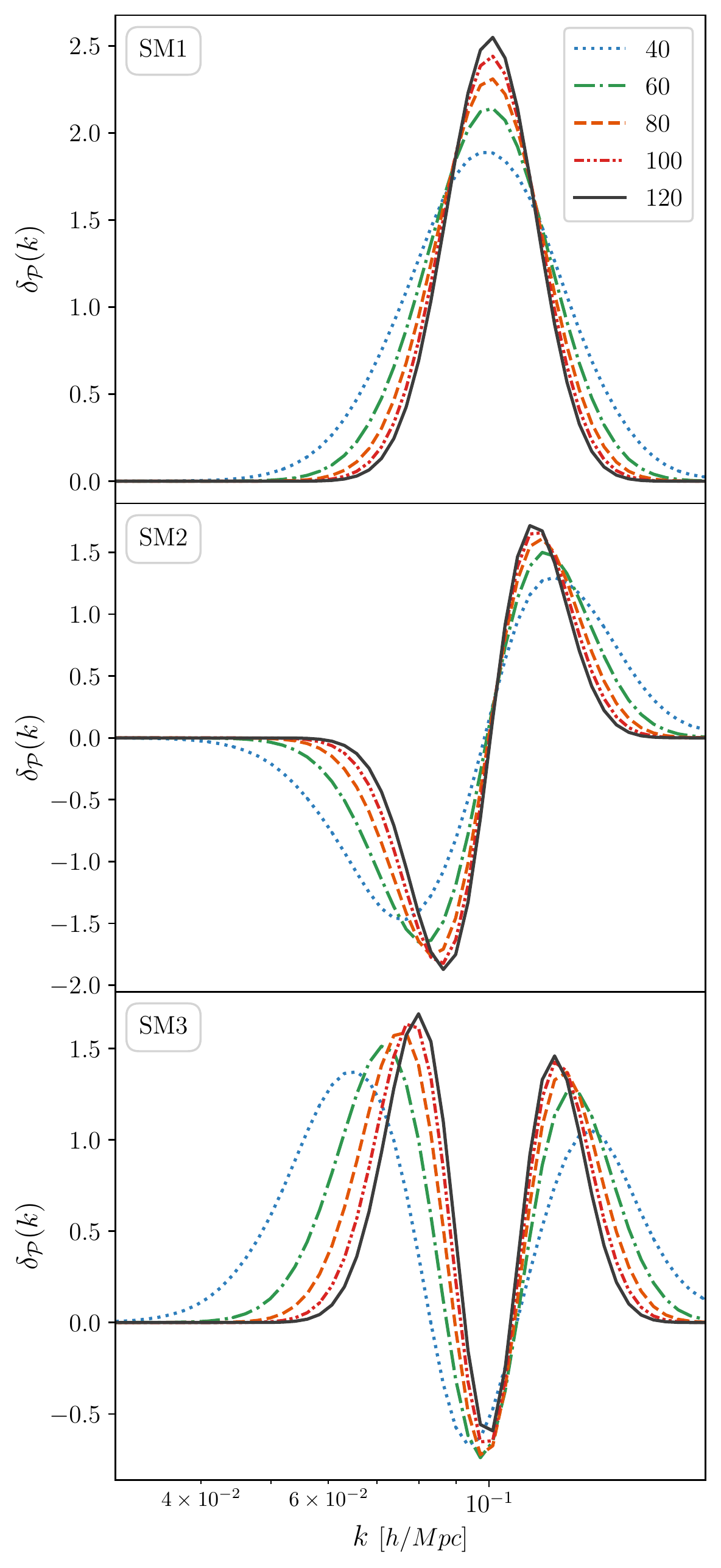}
  \caption{The first three scalar modes, SM1--SM3,  for the GC probe of Euclid, compared for different bin numbers of the basis function (here, Gaussian). Plots for the convergence of the modes for the other WL and CMB probes show similar behaviors. We see that the modes are practically converged for $N=120$.}
    \label{fig:NMds}
  \end{center}
  \end{figure}
    %---------------------------
  The light of distant galaxies is distorted by the gravitational potential of the large scale structures on its way to reach the observer. This distortion is referred to as weak lensing, or WL, and is conveniently characterized by the two parameters, shear  $\gamma$ and the scalar convergence $\kappa$ \citep[see, e.g., ][for a recent review]{Kilbinger_2015}.
The convergence power spectrum of cross correlation of the bin pair $(i,j)$, using the Limber approximation, is given by \citep{kaiser1998weak,Hu:1999}

    \begin{equation}
    \begin{aligned} 
    P^{\kappa}_{i j}(\ell)= & \frac{9}{4} \int_{0}^{\infty} \mathrm{d} z \frac{W_{i}(z) W_{j}(z) H^{3}(z) \Omega_{m}^{2}(z)}{(1+z)^{4}} \\ & \times P_{\rm m}(\ell \slash r(z), z)
    \end{aligned}
    \end{equation}
   where  $\Omega(z)$ represents the density parameter at redshift $z$ and $r(z)$ is the comoving distance from observer at $z=0$ to redshift $z$.   
   The window function of the $i$th bin is described by   

    \begin{equation}
    W_{i}(z) = \int_{z}^{\infty} dz^{\prime} \, \left[1 - \frac{r(z)}{r(z^{\prime})}\right] \, n_{i}(z^{\prime}),
    \end{equation}
  where  \citep{Ma:2006}

     \begin{equation}
    n_{i}(z) \propto \int_{z_{i, {\rm low}}}^{z_{i, {\rm up}}} dz_{\rm ph}\, \frac{dn_{\text{gal}}}{dz}(z)\, p(z_{\rm ph}, z)
    \end{equation}
 with $p(z_{\rm ph}, z)$ representing the photometric redshift distribution, assumed Gaussian in this work. The unnormalized source distribution function $\frac{dn_{\text{gal}}}{dz}$ is modeled as

    \begin{equation}
    \frac{dn_{\text{gal}}}{dz} = z^{\beta} \, \exp \left[- (\frac{z}{\alpha z_{0}})^{\eta}\right].
    \label{UNDR}
    \end{equation}
 Figure~\ref{fig:NSRDF} illustrates this distribution for the Euclid and SKA surveys.
 
 Finally, the covariance matrix of the shear power spectrum for the redshift pair $(i,j)$ is given by 
    \begin{equation}
    C^{\kappa}_{ij} = P_{ij}^{\kappa} + \delta_{ij} n_{j}^{-1} \gamma_{\text{int}}^{2}.
    \end{equation}
    where $\gamma_{\text{int}}$ is the intrinsic galaxy ellipticity and $n_{j}$ represents the shot noise
    \begin{equation}
    n_{j} = 3600 (\frac{180}{\pi})^{2} n_{\theta} \slash \mathcal{N}_{\text{bin}}.
    \end{equation}
See Table~\ref{sp2table} for the experimental specifications of the WL surveys used in this work.
 Given the power spectra and their covariance matrix, the Fisher matrix for the WL can be obtained by summing over all possible correlations at different redshift bins \citep{Eisenstein:1999},

    \begin{equation}
    \begin{aligned}
        F_{\alpha \beta} = & f_{\text {\,sky}}\, \sum_{\ell}\,  \frac{(2 \ell+1) \ell}{2} \Delta \ln \ell \sum_{ijkl}\,  \left[\frac{\partial P_{ij}}{\partial q_{\alpha}}\, C_{jk}^{-1}\, \frac{\partial P_{kl}}{\partial q_{\beta}}\, C_{li}^{-1}\right]  \\
        = &f_{\text {\,sky}}\,  \sum_{\ell}\, \frac{(2 \ell+1) \ell}{2} \Delta \ln\ell 
         \operatorname{Tr} \left[\frac{\partial {\bf P}}{\partial q_{\alpha}}\, {\bf C}^{-1}\, \frac{\partial {\bf P}}{\partial q_{\beta}}\, {\bf C}^{-1}\right] 
        \label{FWL}
    \end{aligned}
    \end{equation}
    where $f_{\text{\,sky}}$ is the fraction of the sky covered by the survey and $i$, $j$, $k$ and $l$ label the redshift bins.
  For the construction of the Fisher matrix with the WL data one should again set a minimum scale for which the power spectrum can be considered trustable and not significantly contaminated by nonlinear theoretical uncertainties. 
  Here we follow  \cite{Sprenger_2019} to set a sharp cutoff on $k$
  \citep[see][for other suggestions of dealing with this theoretical uncertainty]{Casas_2016}.
  We find that, unlike the case for GC, the results of WL are not highly sensitive to small changes in the choice of the cutoff.
  
  %----------------------------
\section{Results}\label{sec:results}
  %---------------------------
 Given the Fisher matrices for the perturbations in the PSPS using the forecasts of the CMB, GC and WL simulations, we can proceed by their eigen-decomposition to find the principal modes of the assumed model-independent perturbations.
 
  %------------------------
  \subsection{Modes}\label{sec:modes}
 %---------------------------
  In this section we present the scalar (eigen)modes, or SMs, constructed with the method discussed in Section~\ref{sec:method} for the various datasets used in this work.
Figure~\ref{fig:NMds} compares the first three eigenmodes for different number of bins, for the GC. 
We see that in all cases the overall shape of the mode is relatively robust to the increase in the mode number from $N\sim40$ to $N\sim 120$ and the positions of the peaks and troughs do not significantly change. However, their widths decrease as the  number of used basis function increases. The modes for the other probes used in this work show  similar convergence behavior and we verified that the first few modes are practically converged for $N=120$ for all cases. Therefore the results  in the rest of this work are based on $N=120$. 
We also verified that the modes look the same for the different local and nonlocal basis functions considered here. 

   %------------------------------------------------------
  \begin{figure}
  \begin{center}
  \includegraphics[scale=0.65]{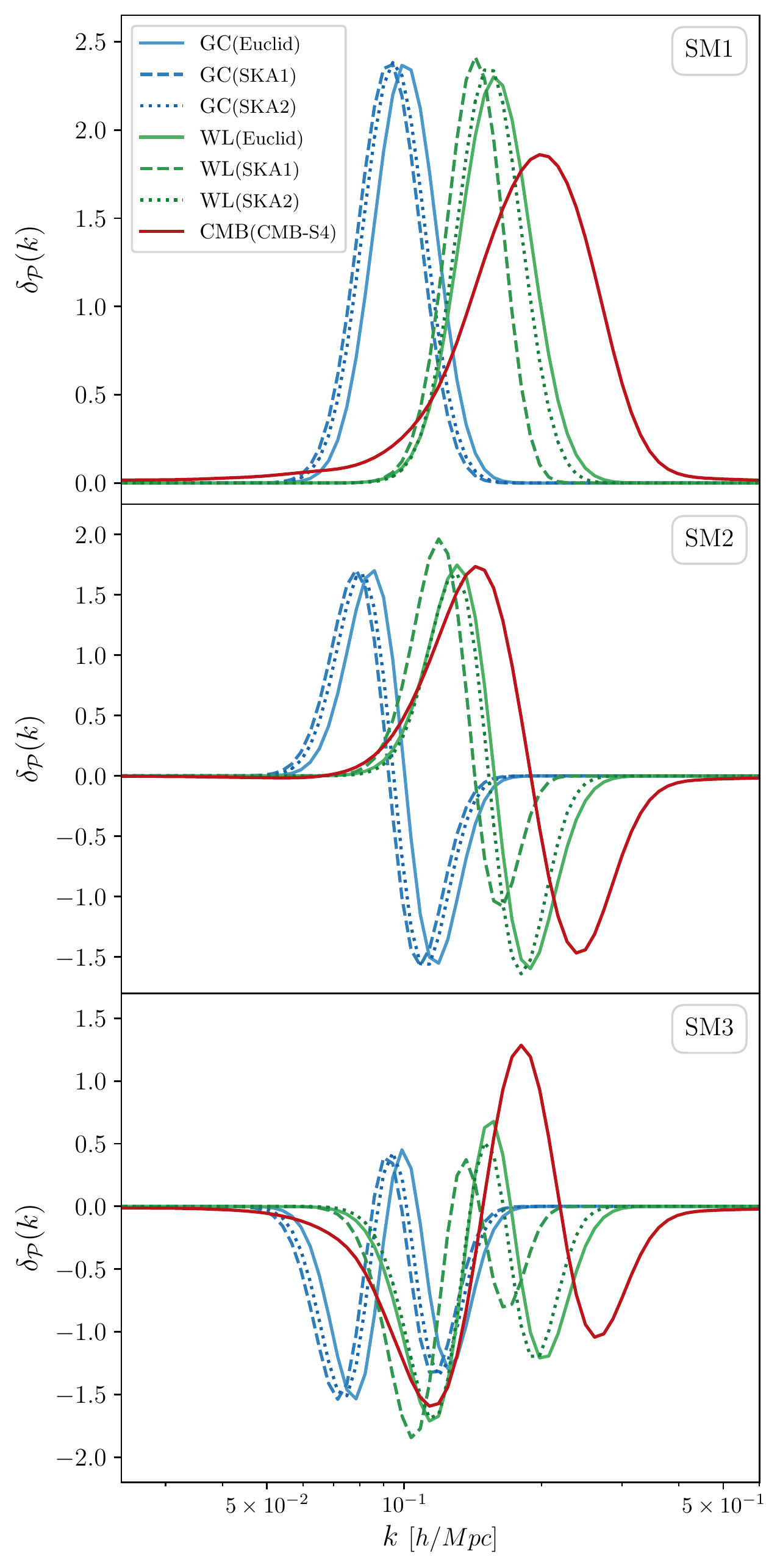}
 \caption{The first three scalar modes, SM1-SM3, constructed for the different future probes considered in this work. The blue, green and red lines coorespond to GC, WL and CMB probes, with different experimental specifications.}
     \label{fig:convergence}
  \end{center}
  \end{figure}
\begin{figure}%[!b]
  \begin{center}
  \includegraphics[scale=0.7]{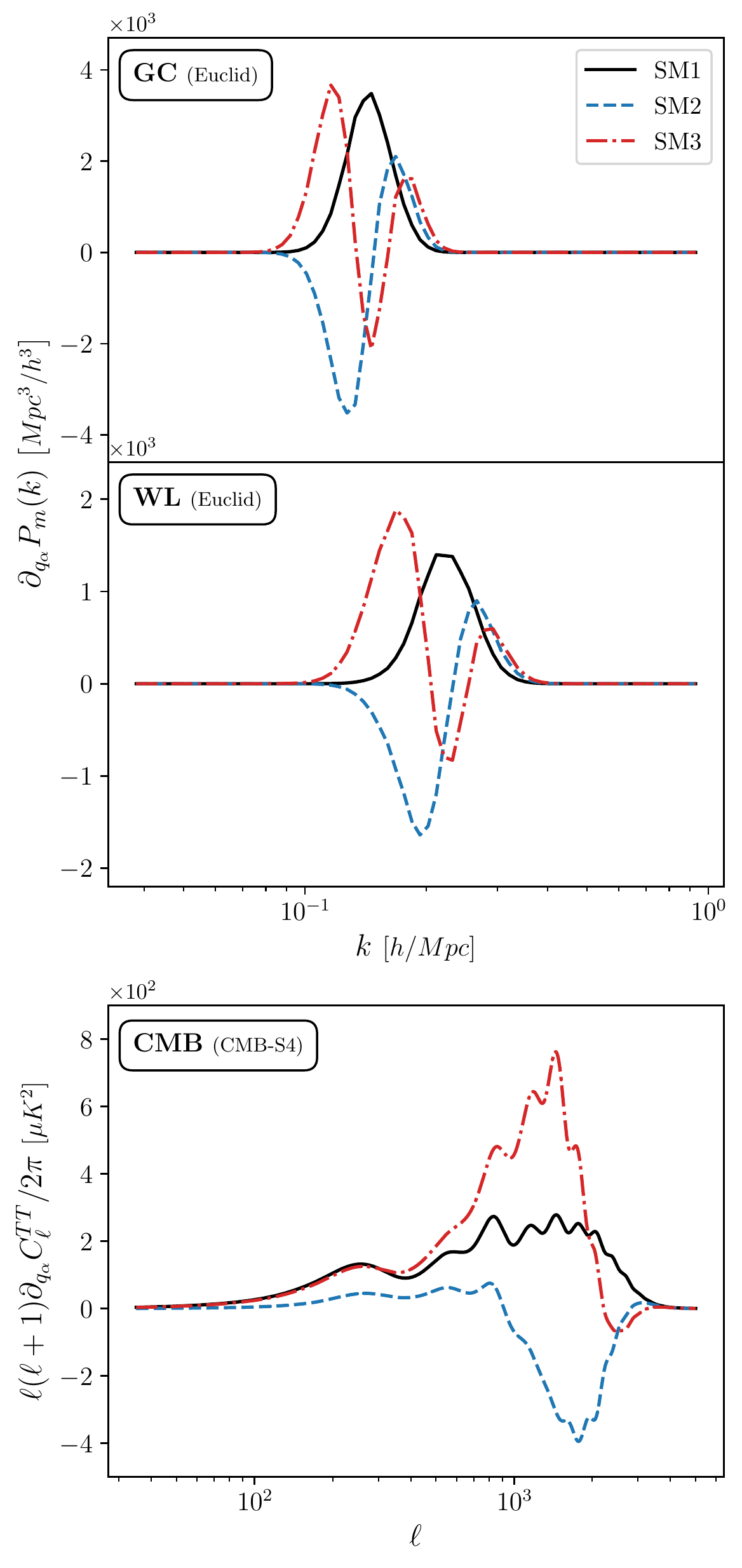}
  \caption{The response of the matter power spectrum to perturbations in the PSPS in the form of the first three  SMs, generated for the Euclid GC and WL probes (top and middle) and CMB--S4 simulations (bottom).}
    \label{fig:sem}
  \end{center}
  \end{figure}

   %-------------------------------------------------------
   Figure~\ref{fig:convergence} compares the first three eigenmodes for the different surveys and Figure~\ref{fig:sem} illustrates the response of the matter and CMB power spectra to changes in the PSPS induced by these modes. The overall shapes of the modes, in terms of the number of bumps and their width and heights are similar (not exactly so for the first CMB-based modes). However, the features seem to be shifted in the $k$-space for the different probes, implying that the different probes are sensitive to different parts of the spectrum.
   Depending on the true possible pattern of spectrum of the sky, any of these could be the one to better reconstruct the pattern. For example, if the true sky is different from the power-law model at larger wavelengths, the GC works best, while for those perturbing the smaller scales, the CMB-S4 would be most sensitive. 
In terms of the estimated errors, we find that the CMB and GC perform in general better than the WL probes. 
The sky coverage impacts the estimated errors of the modes through the factor of $f_{\rm sky}$ in the Fisher matrix. It may also affect the mode shape through changing the minimum wavenumber (multipole for CMB)  included in the Fisher matrix. 
For the first CMB-based mode, however,  we find that the modes look relatively the same and the change in the sky coverage and therefor $\ell_{\rm min}$ does not impact the first few modes. This can be expected as the first mode is centered around $k\sim 0.2-0.3$, while the $\ell_{\rm min}$'s for the two sky  coverages potentially translate into modifications to much smaller $k$'s. 

The modes constructed in this section describe the patterns that data are most sensitive to. If the underlying power spectrum of the sky deviates from the standard $\mathcal{P}_0(k)$ in patterns  similar to these modes, the deviations would be most significantly detectable.
In other words, the projections of the underlying spectrum along these modes are the only part reconstructable by data. To detect patterns which are hugely different from these modes strong theoretical priors are required.
   %-------------------------------------------------------

  %------------------------------------------------------
  \subsection{Number of Modes}\label{sec:Nmodes}
%------------------------------------------------------
The eigen-decomposition of the Fisher matrix generates a mode hierarchy ordered with their estimated errors. The total number of generated eigenmodes is equal to the number of basis functions used in the first place to introduce perturbations in the PSPS. However,  numerical errors render the very high modes practically numerical noise-dominated and should be discarded. 
On the other hand,  all of the first few modes are not necessarily informative enough to be included in further analysis.  
The inclusion of more modes leads to more complexity due to new degrees of freedom, only to explain the very same set of data that could be potentially explained by a lower-dimensional parameter space. Therefore one should impose with care a criterion to truncate this hierarchy and include a limited number of modes so that the enhanced information gained by the modes is worth the increased complexity of the description. 
The estimated errors of the modes can give insight to make the  choice. 

\begin{figure}
  \begin{center}
  \includegraphics[scale=0.5]{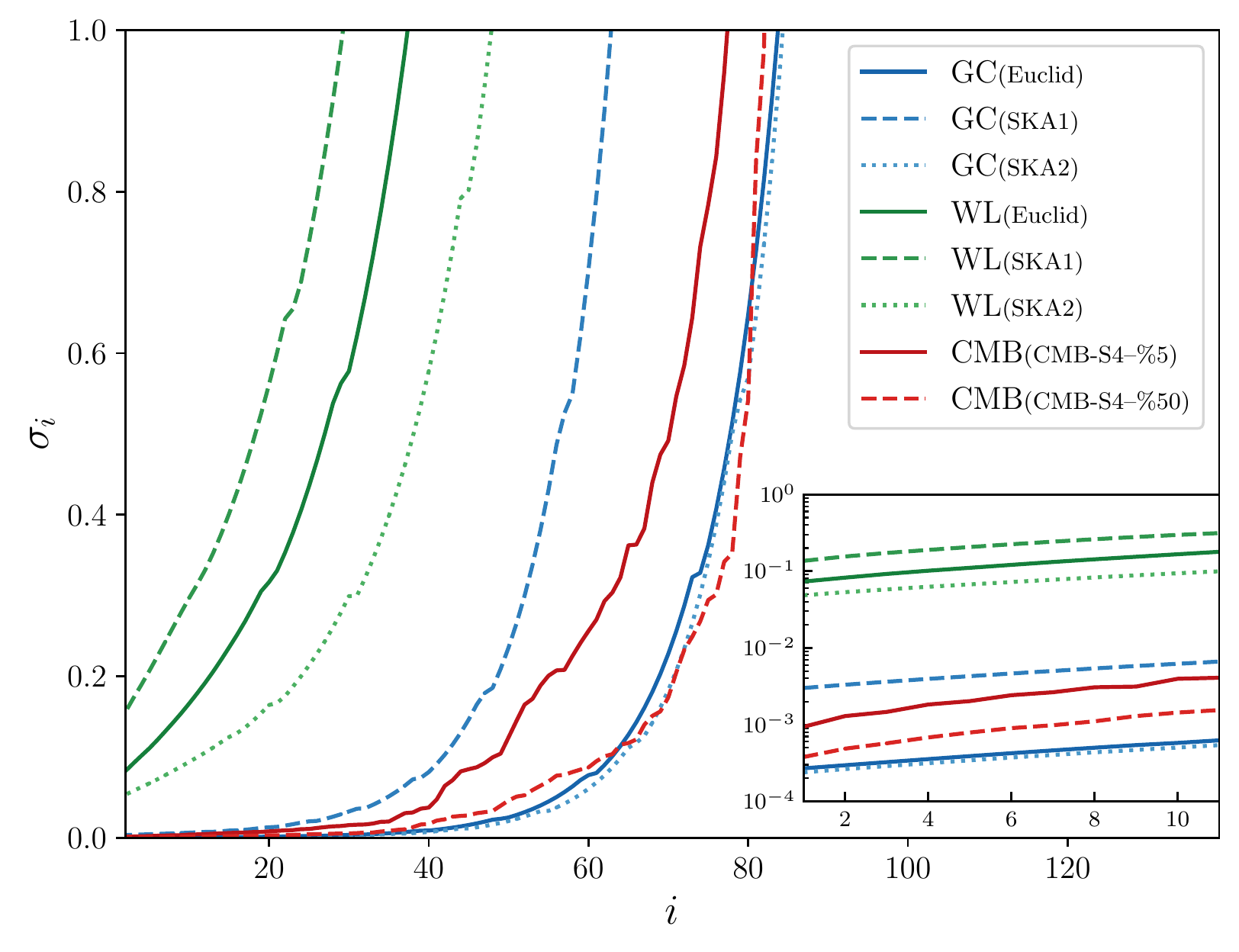}
  \caption{The estimated errors on the eigenmodes of PSPS perturbations,  forecasted by the Fisher analysis for various future surveys. The blue, green and red lines correspond to GC, WL and CMB probes, with different observational  specifications.  }
    \label{fig:sig}
  \end{center}
  \end{figure}

Figure~\ref{fig:sig}
shows the estimated errors of the modes and compares them for different probes.  We see that, e.g., for the WL surveys the increase in errors is sharp even at very low mode numbers, implying the high cost of the introduction of the new modes. For the CMB-based modes, on the other hand, there is a plateau, with comparable errors, for the first few modes. 
We also find that the errors of the WL-based modes are the highest compared to the GC-- and CMB--based modes.  
%------------------------------------------------------
  %------------------------------------------------------
\begin{figure}
\begin{center}
\includegraphics[scale=0.5]{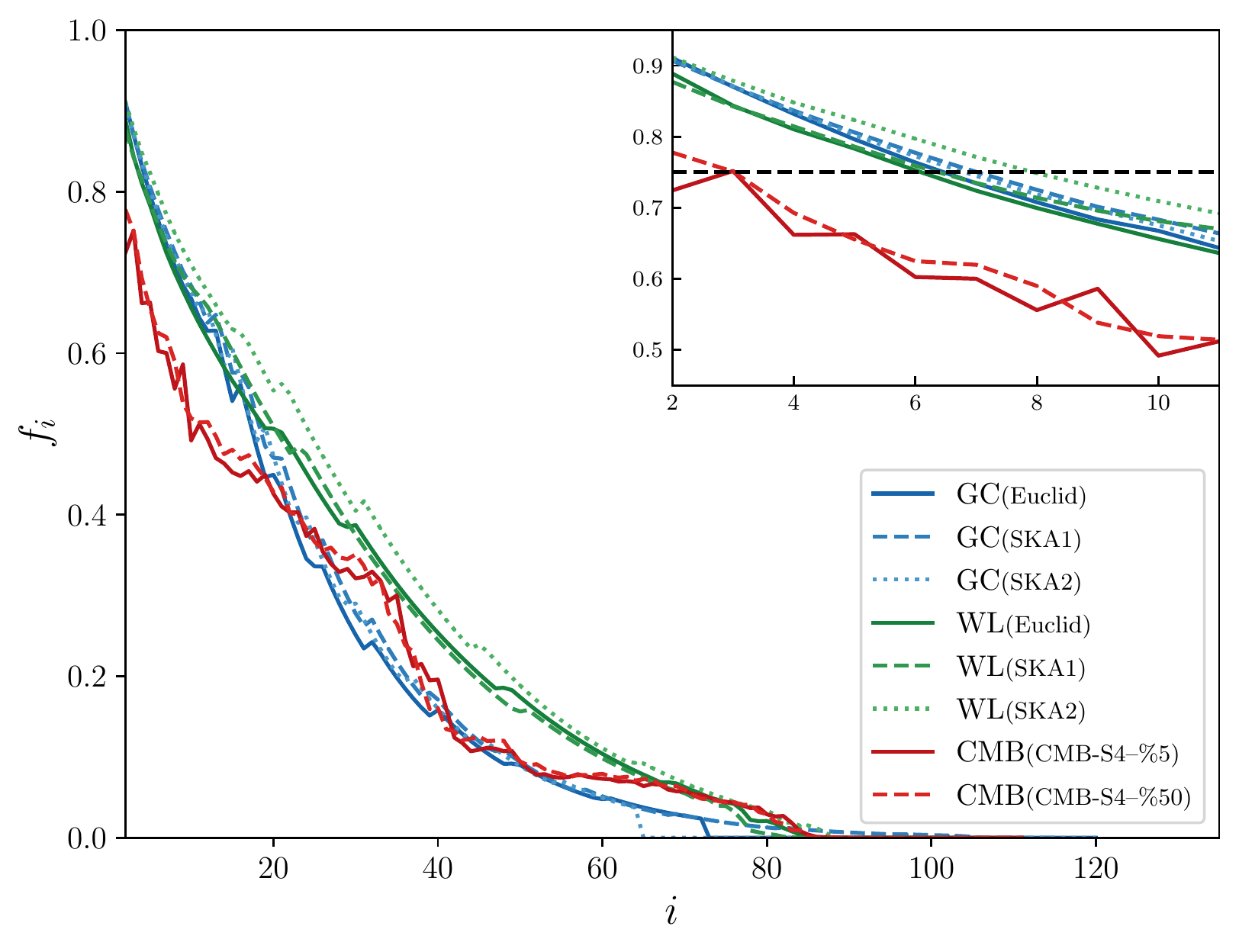}
\caption{Similar to Figure~\ref{fig:sig}, but for the FOM, introduced in Equation~\ref{eq:fom}.}
  \label{fig:fom}
\end{center}
\end{figure}
%
%--------------------------
Here we introduce a figure of merit, or FoM, to quantify the information delivered by a mode compared to the average information delivered by the lower modes with smaller error bars.
The inverse of the volume in the parameter space of the first $(i-1)$ modes spanned by their $1\sigma$ confidence intervals, ${\cal I} \sim (\prod_{j=1}^{i-1} \Lambda_j)^{1/2}$, is a measure of the information learned about the perturbations if the first $(i-1)$ modes are included in the analysis. 
The average contribution of each mode to this information content is ${\cal I}^{1/(i-1)}$. We define the FOM for the inclusion of the $i$th mode, or $f_i$, to be its relative contribution to the total acquired information compared to the average of the previously included modes, 

\begin{equation}
f_i=\sqrt{ \Lambda_i\slash \big( \prod_{j=1}^{i-1} \Lambda_j \big)^{\frac{1}{i-1}}}~.\label{eq:fom}
\end{equation}
Figure~\ref{fig:fom} illustrates the FoMs for the modes from the different probes. The $f$'s are expectedly smaller than $1$ as the modes, with increasing errors, are in decreasing order of impact on the analysis and therefore deliver less information compared to the average of the lower modes. A proper cutoff on the FoM, depending on the experiment and the detailed goals of the analysis, could be chosen to break this hierarchy. For example, a cutoff at $f_{\rm cut}=0.75$ would exclude the modes if their contributions drop below $75\%$ of the average of its previous modes. This would keep the first three  modes for the CMB-S4 and the first six modes for the large scale surveys.  
Alternatively, a cutoff could be set where there is a significant gap, or a sharp decrease, in the FoM.  
However, we do not consider this approach  as  the FoM for the modes produced here decreases quite smoothly with no sharp gap.

%------------------------------------------------------
\subsection{Physically Motivated Models for PSPS}\label{sec:PMMs}
%------------------------------------------------------
Here we investigate how the modes introduced in this work perform in the reconstruction of the power spectra motivated by 
certain models of the early Universe.
In other words, we assume the underlying initial conditions  of the {\it true} sky are described by certain forms of power spectra that deviate from the standard inflationary power-law spectrum $\mathcal{P}_{0}(k)$.
The goal is to explore whether these  deviations from the standard scenario are recoverable by data itself, in the absence of prior biases and theoretical prejudices. We do this by using the eigen modes of Section~\ref{sec:modes} to reconstruct the underlying power spectrum.
To be specific, we use two largely different models for the power spectrum: the Step model  with localized features (Section~\ref{sec:StM}), and  the Logarithmic Oscillations model with oscillations extended in the
full $k$-space (Section~\ref{sec:LOM}).
In this section we use the modes generated based on simulations of the CMB--S4  sky and the Euclid specifications of GC and WL probes. The SKA modes are quite similar to the Euclid case.
  %------------------------------------------------------
  \subsubsection{Step Model}\label{sec:StM}
%------------------------------------------------------
The Step model was initially introduced as an abrupt feature in the inflation model \citep{Adams2001} or in the sound speed \citep{Achucarro2011} which leads to a locally oscillatory behavior in the PSPS. A more recent general parametrization of this model is described in \cite{Miranda2014} and introduces the PSPS as
\begin{equation}\label{eq:s}
\mathcal{P}_{s}(k) = \exp\left[\ln \mathcal{P}_{0}(k)  +\mathcal{I}_{0}(k) + \ln \left(1 + \mathcal{I}_{1}^{2}(k)\right)\right],
\end{equation}
where 
\begin{equation}
\begin{split}
\mathcal{I}_{0}(k) &\approx \mathcal{A}_{s} \mathcal{W}_{1}^{(0)}(k \slash k_{s}) \mathcal{D}\left(\frac{k \slash k_{s}}{x_{s}} \right),\\
\mathcal{I}_{1}(k) &\approx \frac{1}{\sqrt{2}} \times\\
&\left[\frac{\pi}{2}(1-n_{s}) + \mathcal{A}_{s} \mathcal{W}_{1}^{(1)}(k \slash k_{s}) \mathcal{D}\left(\frac{k \slash k_{s}}{x_{s}} \right)\right].
\end{split}
\end{equation}
The window functions  are defined as
\begin{equation}
\begin{split}
\mathcal{W}_{1}^{(0)}(x) &= \frac{1}{2 x^{3}}\\
&\left[\left(18x - 6x^{3}\right)\cos 2x + \left(15x^{2}-9\right)\sin 2x\right],\\
\mathcal{W}_{1}^{(1)}(x) &= \frac{-3\left(x \cos x -\sin x\right)}{x^{3}}\\ 
&\left[3x \cos x + \left(2x^{2}-3\right) \sin x\right],
\end{split}
\end{equation}
and $\mathcal{D}(x) = x \slash \sinh x$ is the damping function. 
The dashed red curve in Figure~\ref{fig:psms} shows the primordial power spectrum for the Step model for the parameter set $(\mathcal{A}_{\rm s}, k_{\rm s}, x_{\rm s})=(0.374, 0.04h/{\rm Mpc}, 1.4)$ consistent with the prior range assumed in \cite{pl18inf}.
  %------------------------------------------------------
  \subsubsection{Logarithmic Oscillations Model}\label{sec:LOM}
\begin{figure}
  \begin{center}
  \includegraphics[scale=0.66]{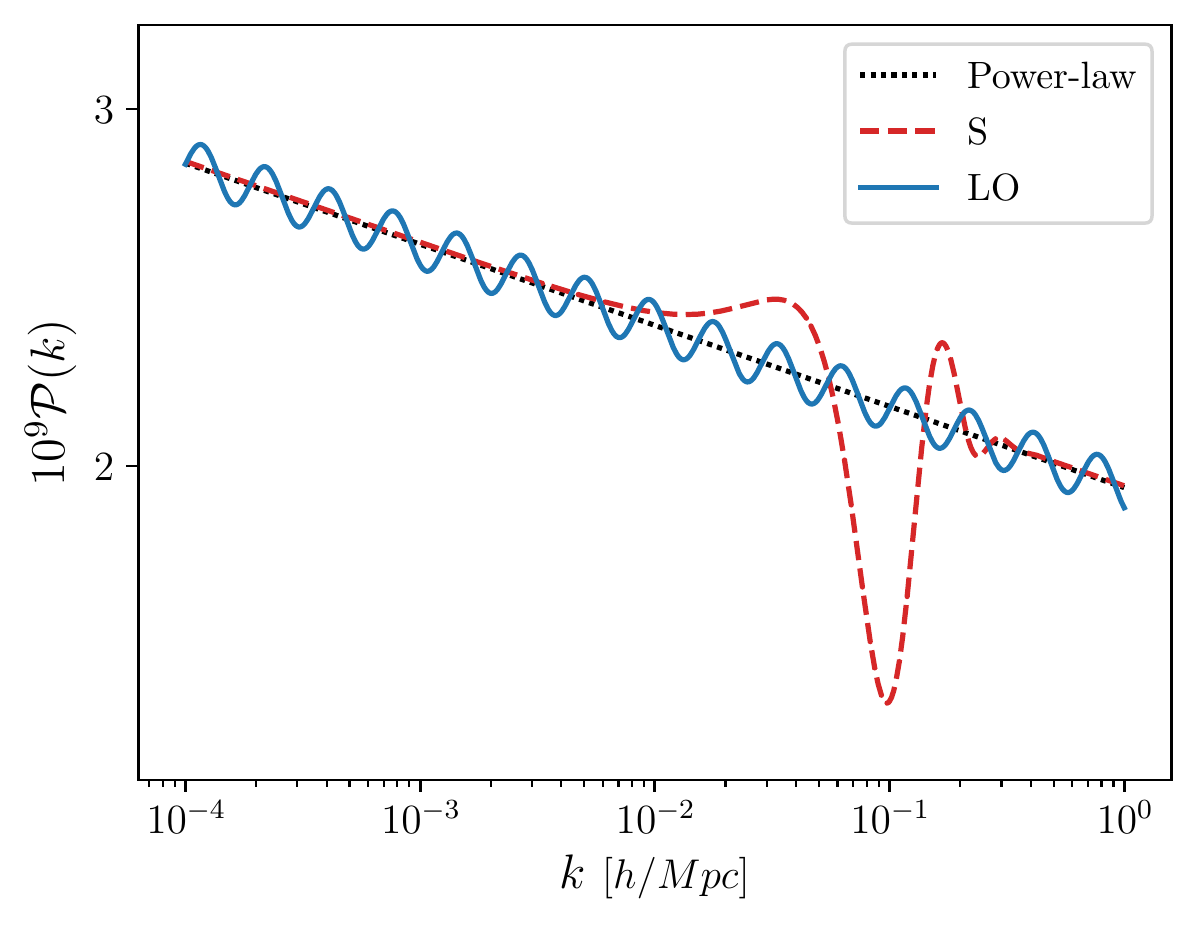}
  \caption{The primordial  power spectrum for $\mathcal{P}_{0}(k)$ (the power-law model), the Step (S) and the Logarithmic Oscillations (LO) models.}
    \label{fig:psms}
  \end{center}
  \end{figure}

%------------------------------------------------------
As the second case we consider an oscillatory deviation from the standard PSPS,  extended through the observed $k$-range as \citep[see, e.g.,][]{pl18inf},
\begin{equation}\label{eq:lo}
\begin{split}
\mathcal{P}_{\log}(k) &= \mathcal{P}_{0}(k) \times\\
& \left[1 + \mathcal{A}_{\log} \cos \left(\omega_{\log} \ln(\frac{k}{k_*}) + \phi_{\log}\right)\right].
\end{split}
\end{equation}
This type of power spectrum appears in models with non-Bunch-Davis initial conditions \citep{Martin2001,Danielsson2002,Bozza2003} and in axion-monodromy models \citep{Silverstein2008,McAllister2010,Kaloper2011,Flauger_2017}. %
The dotted blue curve of  Figure~\ref{fig:psms} illustrates the power spectrum for this model with the parameter set $(\mathcal{A}_{\rm log},\omega_{\rm log}, \phi_{\rm log})=(0.0278,10 ,2\pi(0.634))$ consistent with the prior range of \cite{pl18inf}.
%------------------------------------------------------
%------------------------------------------------------
  \subsubsection{Reconstruction}\label{sec:PMRec}
%------------------------------------------------------
%------------------------------------------------------
  \begin{figure*}
  \begin{center}
  \includegraphics[scale=0.55]{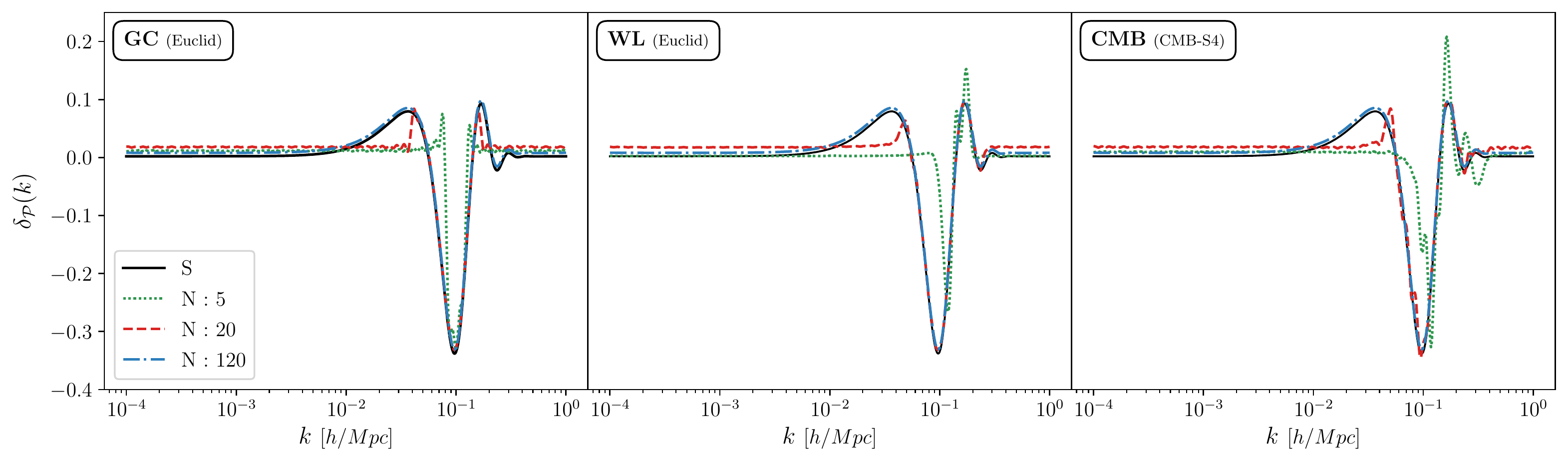}
  \caption{The reconstructed Step model with  5, 20 and 120 modes (the dotted green, dashed red and dash-dotted blue respectively) compared to the full underlying pattern (solid black) labeled as "S". The different plots correspond to reconstruction with different simulated datasets.}
    \label{fig:stcm}
  \end{center}
  \end{figure*}
  %------------------------------------------------------
  %------------------------------------------------------
  \begin{figure*}
  \begin{center}
  \includegraphics[scale=0.55]{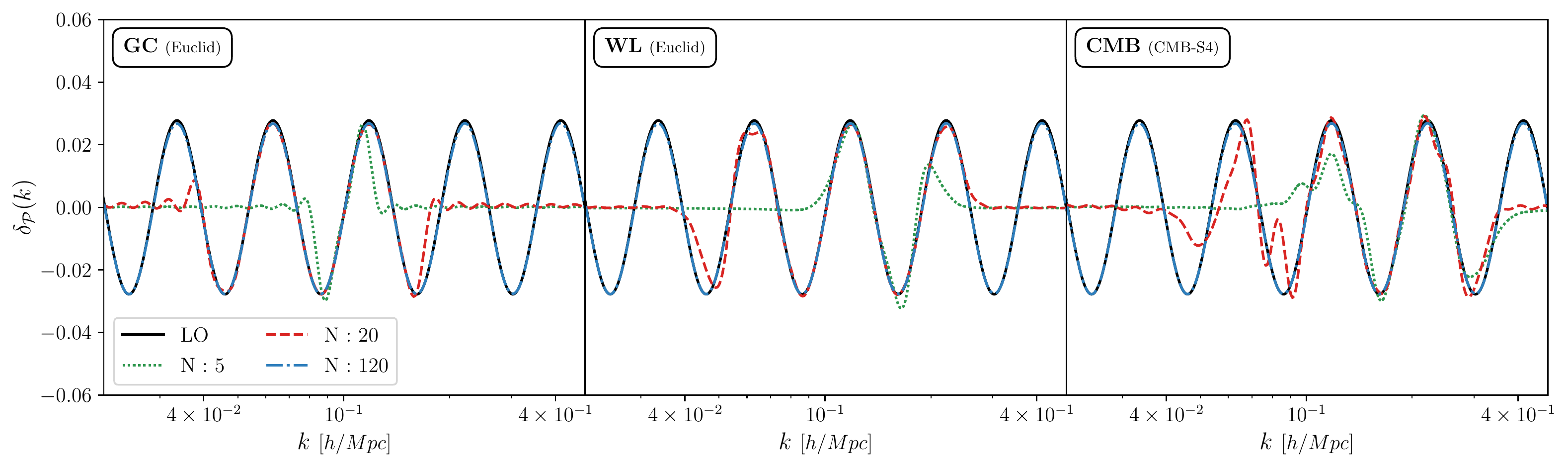}
  \caption{Similar to Figure~\ref{fig:stcm} but for the Logarithmic Oscillations model labeled as LO.}
    \label{fig:locm}
  \end{center}
  \end{figure*}
%------------------------------------------------------
Given a model of the early Universe, one can make forecast on the detectability of the features it predicts for the primordial power spectrum. 
The analysis pipeline developed in the previous sections is  not biased toward certain theoretical 
predictions and allows for data-driven degrees of freedom in the parameter space. In principle, these new parameters allow for the reconstruction of the underlying initial conditions of the structures in the sky. In practice, however, the model prediction for the PSPS could be hugely different from the patterns distinguished by data as the most observable. In such cases a large number of modes from the mode hierarchy are required for an acceptable reconstruction of the underlying spectrum.

In this section we explore the recoverability of the underlying PSPS, $\mathcal{P}(k)$, of Equations~\ref{eq:s} and \ref{eq:lo} by the modes $\mathcal{E}_i(k)$ constructed for the simulations of CMB, GC and WL. 
One can expand the $\mathcal{P}(k)$ as 
\begin{equation}
\mathcal{P}(k) \approx \sum_{i=1}^{N} \alpha_{i} \mathcal{E}_{i}(k)
\label{eq:frec},
\end{equation} 
where the $\alpha_{i}$'s are the projection coefficients of the power spectrum on the modes, 
\begin{equation}
\alpha_{i} = \int_{k_{\text{min}}}^{k_{\text{max}}} \mathcal{E}_{i}(k) \mathcal{P}(k) {\rm d}\ln k.
\end{equation}

We verified the robustness of the reconstruction pipeline and the orthogonality of the modes by using the full set of the $120$ modes and found the close agreement of the recovered power spectrum with the assumed underlying pattern. 
For this purpose we used the Fourier-based modes due to their exact orthogonality.  One could as well use modes based on localized basis functions, remembering that the non-negligible overlap of the neighboring bins could lead to small distortions in the reconstructed power spectrum.
It should however be noted that these numerical imperfections mainly impact modes with relatively large forecasted errors while the first few modes turn out to look very similar for different sets of basis functions.

Figures~\ref{fig:stcm} and ~\ref{fig:locm} compare the reconstructed perturbation to the power spectrum for the two models of Section~\ref{sec:StM} and \ref{sec:LOM}, with different number of modes included in the reconstruction. 
As claimed, we find that the full set of modes makes an elegant  reconstruction of the underlying spectrum in both cases.
For the localized Step model the first five modes can locate the main feature, i.e, the valley at $k \sim 0.1 h/$Mpc, and the small bump to its right. However, many more modes are required to mimic the other bump at lower $k$. This is to be expected, noting that the data is most sensitive to smaller scales ($k \sim 0.1 h/$Mpc), also apparent from Figure~\ref{fig:convergence}. Therefore the best constrainable patterns of deviations, i.e., the lowest modes, are quite flat at larger scales and cannot reproduce features at those scales.
The Logarithmic Oscillations model is much harder to reconstruct due to its extension in a large $k$--range, where the data is hardly sensitive to. We find that the first five modes can generate the oscillatory feature only around the most detectable $k$, i.e., $k\sim 0.1 h/$Mpc. Adding more modes would generate more wiggles at other wavenumbers. 

\begin{figure}
  \begin{center}
  \includegraphics[scale=0.6]{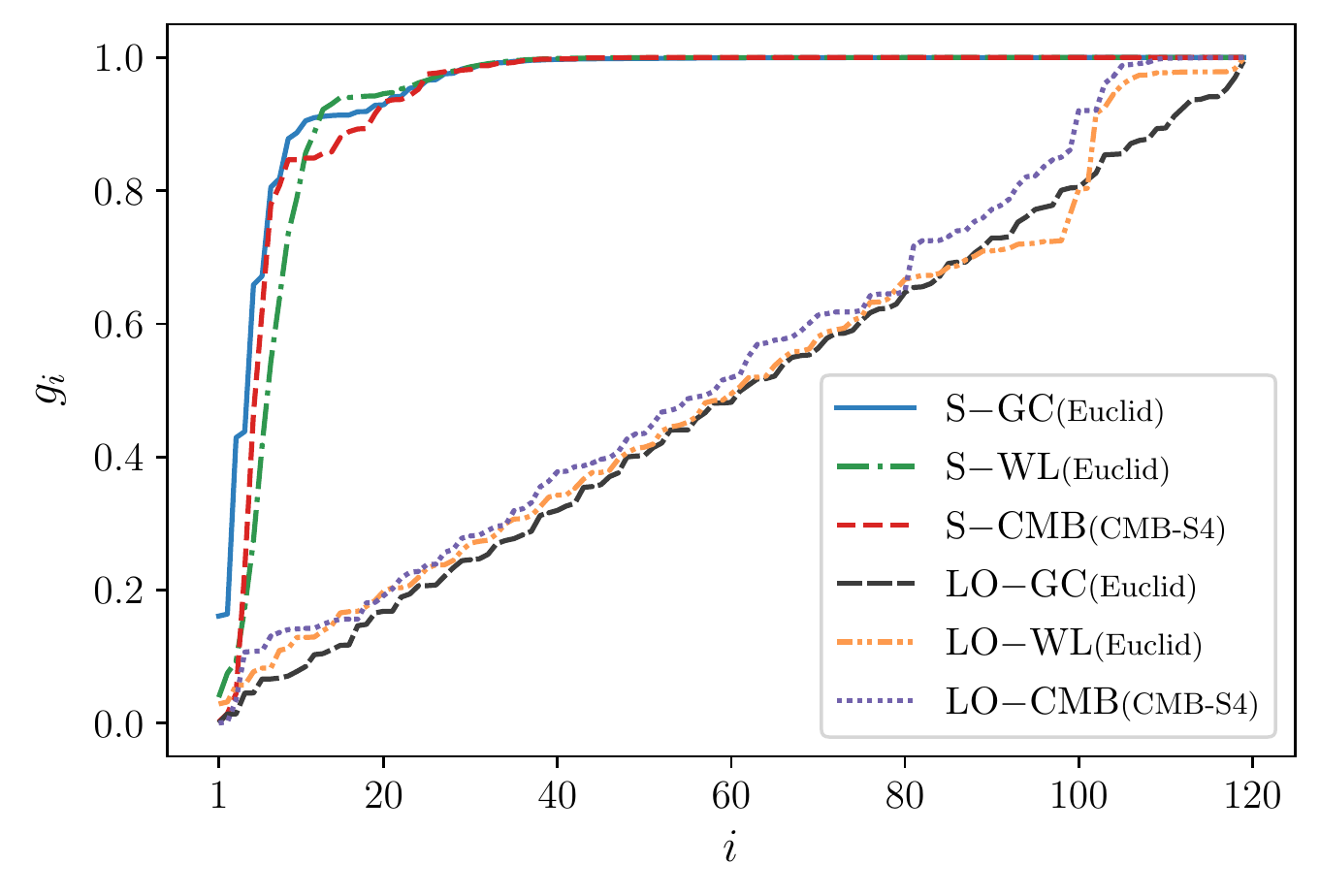}
  \caption{The reconstruction gain $g_i$(see Equation~\ref{eq:gain}) for different number of modes included in the analysis for  the Step and Logarithmic Oscillations models, labeled as "S" and "LO" respectively, and with different simulated datasets.}
    \label{fig:gi}
  \end{center}
  \end{figure}

It should be noted that  reconstruction with higher modes is at the cost of including noisy features in the analysis, and increasing the complexity of the description with more degrees of freedom.
One could get an idea of the number of modes required for a fair reconstruction of a given model by quantifying the relative amount of information delivered by various number of modes compared to the full reconstruction. As a measure we define the {\it reconstruction gain} $g_{i}$
\begin{equation} \label{eq:gain}
g_{i} = \frac{1}{\sum_{j=1}^{N} \alpha_{j}^2}\sum_{j=1}^{i} \alpha_{j}^2 
\end{equation}
where $i$ is the mode number with the highest ranking, i.e., with the largest forecasted error, in the reconstruction, and we assume all the modes with lower errors are included. 
Figure~\ref{fig:gi} shows the gain for the two Step and Logarithmic Oscillations models with different datasets used for the reconstruction. 
For the Logarithmic Oscillations case there is a relatively steady increase as more modes are included.
This is expected as with the extended oscillatory behavior of the spectrum higher modes continue to probe scales which are required for the full reconstruction.
 The reconstruction of the Step model, on the other hand, is quite insensitive to the increase in mode number after $20-30$ modes, as these modes suffice to reproduce the main local features of the pattern.

It is also important to note that despite the increase in the gain, the quality of the reconstruction does not necessarily improve with the increase in the maximum number of included modes. 
That is due to the large uncertainty in the measurement of the higher modes. 
For example we find that in the reconstruction of the Step model with the CMB-S4 simulation, the projected amplitude of the power spectrum gets smaller than the forecasted error of the modes for $i\gtrapprox20$. Therefore, for this model as the underlying power spectrum, the amplitudes of these modes will be measured to be consistent with zero.

%------------------------------------------------------
%------------------------------------------------------

%----------------------------------------------------
\section{Summary and Discussion}\label{sec:discussion}
%----------------------------------------------------
In this work we performed a Fisher--based analysis to search for the observable features in the future large scale and CMB datasets in a model-independent and data-driven way. We found the features that are mostly constrained by data through the eigen--decomposition of the Fisher matrix.
The first few modes were mainly localized in the $k$--range of $(0.05,0.5)$ in units of $h/{\rm Mpc}$. 
The modes looked similar for different probes, shifted with respect to each other in certain cases. 
The main observed pattern in the modes, as the mode number increases, is the increase in the number of peaks and troughs. This can be understood as higher modes have larger estimated errors. More ups and downs in the modes, on the other hand, leads to more ups and downs in the matter power spectrum, which would partially cancel out when summed over all observed scales in the Fisher calculations. 

The Fisher analysis provided a hierarchy of modes which should be truncated according to some criterion. We proposed  an FoM which compares the information content of each mode to the average information-content of lower modes. The mode is accepted only if its FoM is above a certain threshold.  
This FoM is set prior to measurement of the mode amplitudes and is therefore  unbiased by observations, in contrast to, e.g., the criteria of comparing the observed mode amplitude to its estimated error. 

When observations are available, the amplitudes of the modes can be treated  as free parameters alongside the other cosmological parameters, and the measured amplitudes will then be used to reconstruct any perturbations to the primordial power spectrum. If there are hints to certain features in future observations, the modes can be re-generated iteratively with that feature included in the fiducial in the process of mode construction. The current parametrization assigns uniform weight to the bins, which are logarithmically spaced in $k$. This is motivated by the scale-invariance of the PSPS.  One could  consider assigning different weights if there are theoretical motivations or observational hints to search for certain features in different scales.    

%\newpage
\bibliography{seimore}
\bibliographystyle{apj}

\end{document}